\documentclass[preprint,aps,,preprintnumbers,amsmath,amssymb,superscriptaddress,floatfix]{revtex4}

\usepackage{graphicx}
\usepackage{color}
\usepackage{dcolumn}
\usepackage{bm}

\begin{document}

\title{Tagged Neutron, Anti-neutron and $K^0_L$ beams in an Upgraded MIPP Spectrometer\\}
\preprint{MIPP-NOTE-130}
\author{Rajendran Raja}
\affiliation{
Fermi National Accelerator laboratory\\
Batavia, IL 60510}

\date{\today}

\begin{abstract} 
The MIPP experiment operating with an upgraded data acquisition system
will be capable of acquiring data at the rate of 3000 events per
second. Currently we are limited to a rate of ~30~Hz  due to
the bottlenecks in the data acquisition electronics of the Time
Projection Chamber (TPC). With the speeded up DAQ, MIPP will be
capable of acquiring data at the rate of $\approx$5~million events per
day. This assumes a conservative beam duty cycle of 4~sec spill every 2
minutes with a 42\% downtime for main injector beam manipulations for
the $\bar{p}$ source. We show that such a setup is capable of producing tagged
neutron, anti-neutron and $K^0_L$ beams that are produced in the MIPP cryogenic
hydrogen target using proton, anti-proton and $K^{\pm}$ beams. These tagged 
beams can be used to study
calorimeter responses for use in studies involving the Particle Flow
Algorithm (PFA). The energy of these tagged beams will be known to better 
than $2\%$ on a particle by particle level by means of constrained fitting. 
We expect a tagged beam rate in the tens of thousands a day. The MIPP 
spectrometer thus offers a unique opportunity to study the response of 
calorimeters to neutral particles.

\end{abstract}
\maketitle

\section{Introduction} 

\begin{figure}[h] 
\begin{center}
\includegraphics[width=\textwidth]{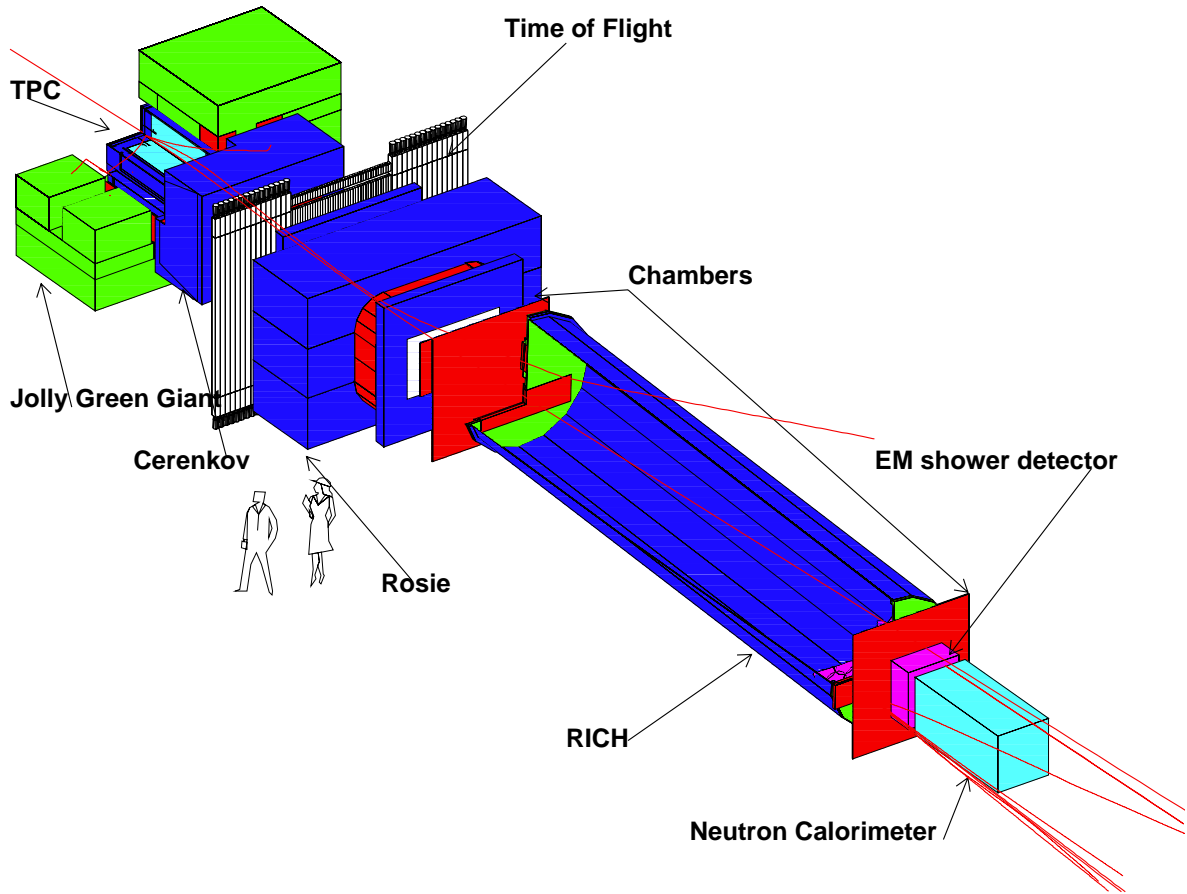} 
\caption{A Geant3 based view of the MIPP Spectrometer. 
The electromagnetic and neutron calorimeters can be replaced by an ILC
test calorimeter to study the response to tagged neutron and $K^0_L$
beams with an upgraded MIPP spectrometer~\label{mipp}}
\end{center} 
\end{figure}

The MIPP experiment at Fermilab is an open geometry spectrometer~\cite{mipprop}
designed to study non-perturbative QCD interactions on a variety of
nuclear targets including liquid hydrogen. It has just completed its
first run using a time projection chamber (TPC) that currently takes
data at $\approx$ 30Hz. Beams of $\pi^{\pm},K^{\pm}$ and $p^{\pm}$ from
a momentum range 5~GeV/c-85~GeV/c have been obtained using the
secondary beamline designed by MIPP. The experiment  has very nearly complete
acceptance of all forward going charged particles and particle
identification is performed using $dE/dx$ (in the TPC), time of
flight, multi-cell Cerenkov an a RICH counter that provides $3\sigma $
separation between $\pi,K$ and $p$ hypotheses over nearly all of
accepted phase space. A schematic of the spectrometer is shown in
 Figure~\ref{mipp}. 

With an upgraded data acquisition system, 
MIPP can be made to take data at 3000Hz. 
Details of the MIPP upgrade which is estimated to cost 
less than \$500,000 may be found in reference~\cite{upgr}.
\subsection{The Beamline}
\begin{figure}[h] 
\begin{center}
\includegraphics[width=\textwidth,angle=-90]{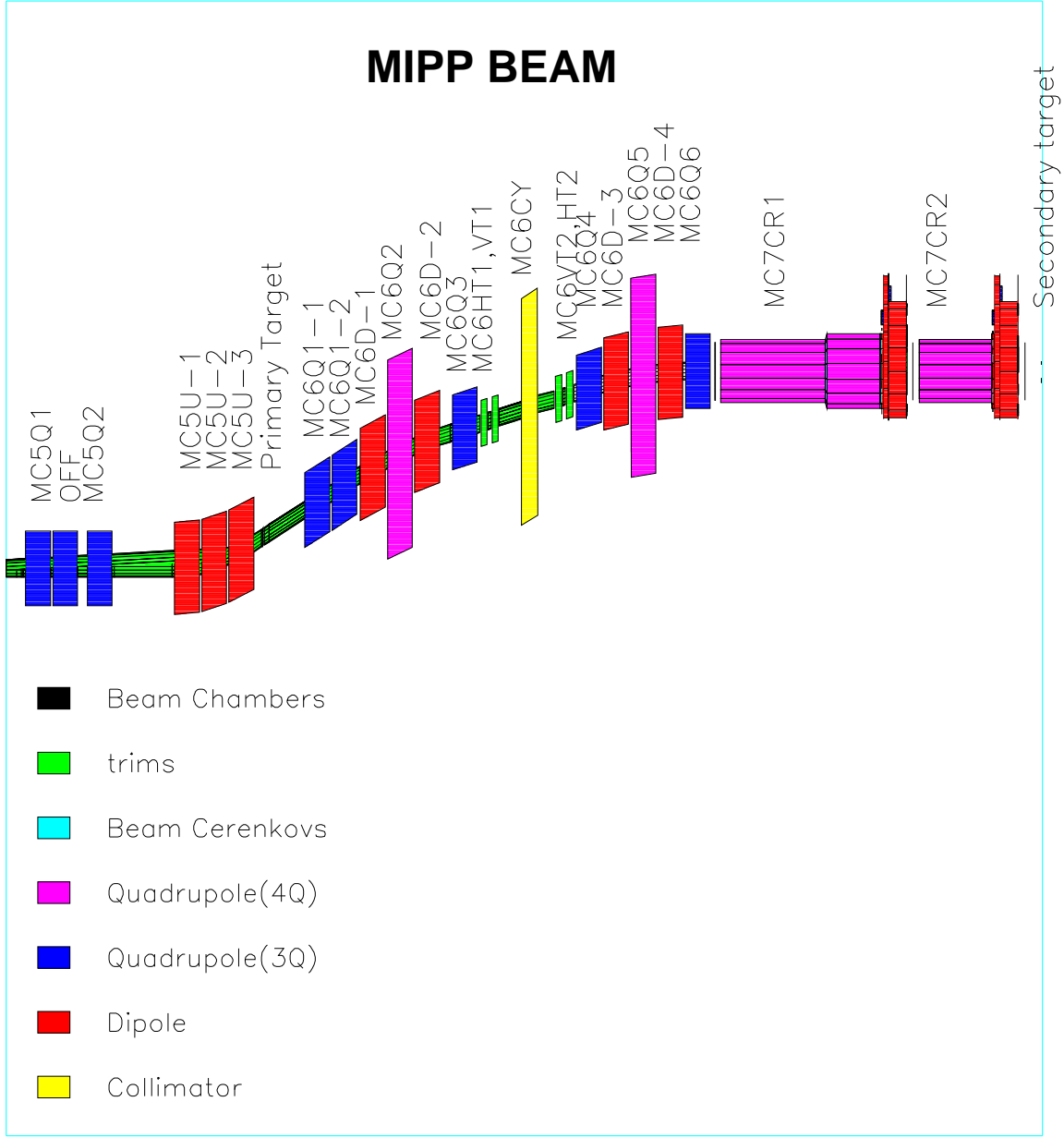} 
\caption{The MIPP secondary beamline. ~\label{beam}
}
\end{center}
\end{figure}

The secondary beamline is shown in Figure~\ref{beam}. The beamline in
an upgraded mode will be capable of delivering charged kaons down to
3~GeV/c and charged pions protons and anti-protons down to
1~GeV/c. 
This excellent performance of the beam is due to the
relatively short distance of 90~meters from the primary production
target to the secondary target. 
The other constraints on the beam are
that the primary target be focused on the momentum selection
collimator which should be in an area of dispersion so that the
momentum bite $\delta p/p$ of the beam can be controlled by opening
the collimator. The collimator is usually set so that $\delta
p/p=0.02$. The other constraint on the beamline is that the divergence
of the beam is small ($<0.3~mr)$ in the region of the beam
Cerenkovs. The MIPP beamline design was selected from a set of 6 different
designs after a great deal of design activity. More details of the
beam may be found in reference~\cite{beam}.  
The beam particle is identified 
by two differential Cerenkovs for particles above $\approx$ 
10~GeV/c momentum and by a time of flight system for particles below this.
The particle identification trigger is ``{\it anded}''
 with the minimum bias interaction trigger 
to provide the experimental trigger. The composition of the beam depends on 
the beam momentum. In order to obtain equal number of interactions for 
all three particle species, the beam triggers are prescaled by the 
appropriate amounts that depend on the beam momentum.
\subsection{Cryogenic Target}
\begin{figure}[h] 
\begin{center}
\includegraphics[width=\textwidth]{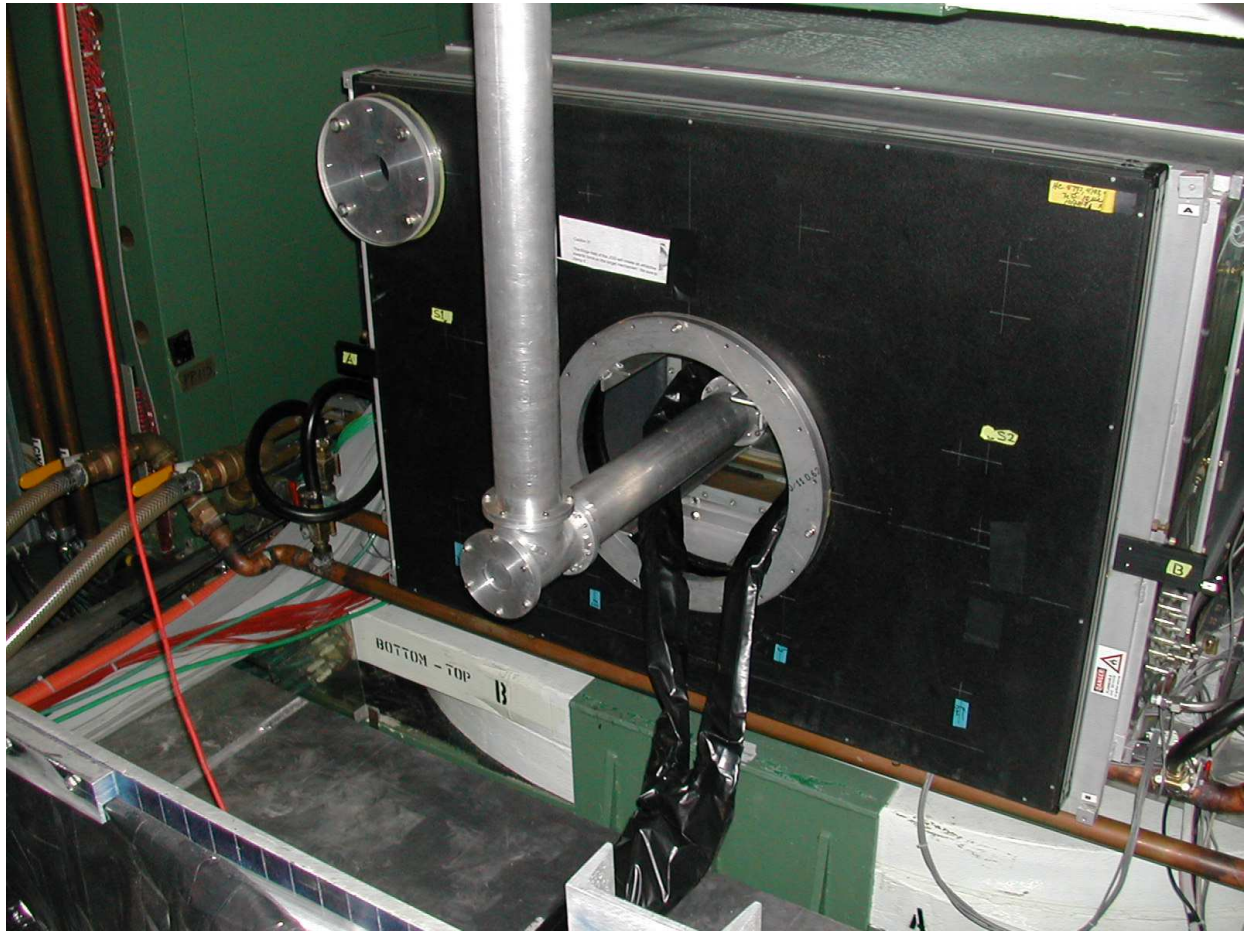} 
\caption{The MIPP cryogenic target filled with liquid hydrogen operating
 in the TPC bay during the first MIPP run.~\label{cryo}
}
\end{center} 
\end{figure}

The tagged neutron, anti-neutron and $K^0_L$ beams require 
the use of constrained
fitting, which make use of the energy momentum constraints at the
primary vertex. This requires that the scattering take place on a
proton target. Nuclear targets will produce many un-detected neutrals
which make constrained fitting next to impossible.  The physics being
proposed here can thus only be done using a liquid hydrogen target. In
the first MIPP run, we operated a cryogenic target that functioned very
well. Figure~\ref{cryo} shows the target installed and operating in the
TPC bay. For a summary of the physics channels available using a
hydrogen target, please see ref~\cite{rajah}.

\section{Diffractive Reactions}

The physics that makes tagged neutral beams possible can be sumamrized in one 
sentence-- Diffractive beam fragmentation in a hydrogen target.

\begin{figure}[h] 
\begin{center}
\includegraphics[width=\textwidth]{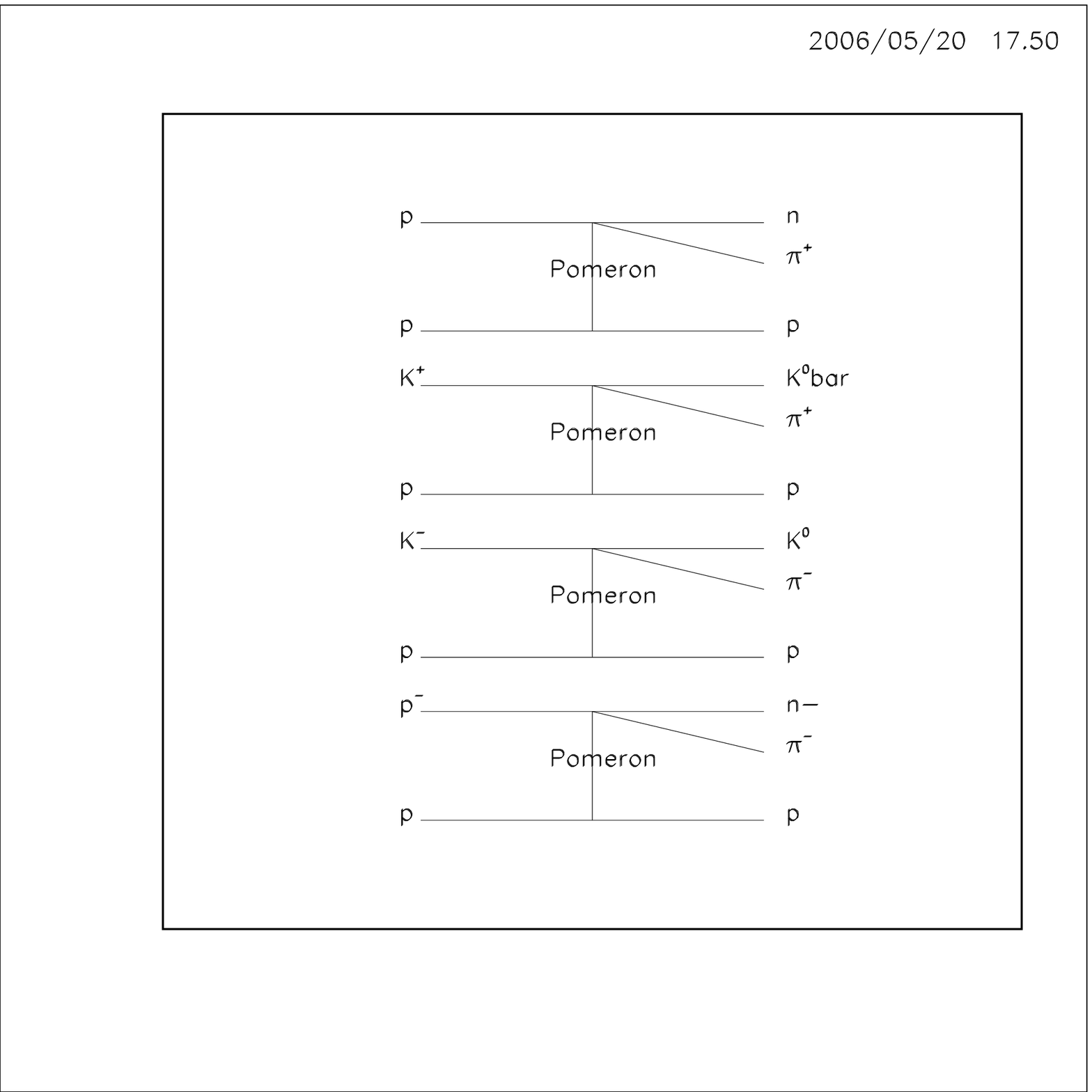} 
\caption{The diffractive diagrams illustrating beam fragmentation for
proton, kaon and anti-proton beams. A pomeron is exchanged causing the beam to
fragment. In the proton case, there is a symmetric target
fragmentation diagram which will cause the emission of slow neutrons
and pions from the target with the beam remaining intact. These
neutrons will be lost~\label{exch}. }
\end{center} 
\end{figure}

The reactions in question are
\begin{eqnarray}
 pp\rightarrow n\pi^+p \\
 K^+p\rightarrow\bar{K^0}\pi^+p ; \bar{K^0}\rightarrow K^0_L\\
 K^-p\rightarrow{K^0}\pi^-p ; {K^0}\rightarrow K^0_L\\
 \bar{p}p\rightarrow \bar{n}\pi^-p 
\end{eqnarray}
These reactions can be picturized by the exchange diagrams shown in 
figure~\ref{exch}.
Diffractive reactions, being due to Pomeron exchange fall relatively slowly 
with incident beam momentum as opposed to pure charge exchange reactions.
\subsection{Constrained Fitting}
The art of constrained fitting was used extensively during the hey-dey
of the bubble chamber whereby it was used to separate various
exclusive reactions from each other. The program SQUAW is an example
of many such computer programs that minimized a $\chi^2$ to obtain
fitted four-vectors of the particles in the reaction allowing for the
measurement errors and correlations.  In a reaction where the beam
momentum 4-vector is assumed known (MIPP identifies the beam particle using beam Cerenkovs and
the beam track is measured using the beam chambers. 
The beam momentum is known to $\approx
2\%$), and the target particle can be reliably assummed to be a
proton, and all the final state particles are identified, one can
apply the conservation of energy and momentum to produce a fit with 4
constraints. Such a fit is known in the jargon as a 4-C fit.
If the final state momenta are measured accurately, then reactions
with missing neutrals will in general tend not to fit. When one of the
final state particles is a missing neutral, the three momentum of the
missing neutral is unknown. The fitting hypothesis assigns a mass to
the missing neutral, and this produces a 1-C fit. For the reactions we
study here, a missing neutron,${\bar n}$ or $K^0_L$ will thus result in a 1-C
fit. If the neutral is observed in the calorimeter and its point of
impact is measured to some accuracy (e.g 
a measurment error in neutral position of 20~cm transversely will 
result in an error in neutral direction of 8mr, with the 
calorimeter at its present position ), one obtains a 3-C
fit. Such 3-C fits are constrained enough to reject events with
further missing $\pi^0$'s, resulting in a fitted momentum of the
$n,{\bar n}$  or $K^0_L$ that is known to $\approx~ 2\%$. This
fitted momentum can then be compared to the measured momentum in the
calorimeter to study the calorimeter response and linearity. The
transverse size of the neutral shower can be studied as a function of
the neutral particle momentum. 

It is worth noting that the quality of the 3-C fit improves 
as the direction of the missing neutral is known better. 
For this, placing the calorimeter farther away from the cryotarget 
improves the precision of the missing neutral direction. Since the 
neutral beam is the result of beam diffraction, the loss in acceptance 
is much less than  
the inverse square of the distance from the interaction point.

\section{Monte Carlo Generation}
The program DPMJET was used~\cite{dpm} to generate the reactions
$pp\rightarrow X$, $K^+p\rightarrow X$, $K^-p\rightarrow X$ and ${\bar p}p\rightarrow X$. We
generated samples of $10^5$ events for $pp$  
interactions at beam
momenta of 10,20,30,60 and 90 GeV/c. Even though MIPP has run with
$K^-$ beams of 85~GeV/c, the prescale factors are large. $K^+$ beams
above 60~$GeV/c$ become problematic, because the proton flux is
large. So we do not generate the 90~GeV/c point for kaon beams. 
For the kaon beams, we generate 200,000 events each, 
since the relative DPMJET cross section seems lower for diffraction 
in $K^{\pm}p$ events. We have generated 100,000 events for 
${\bar p}p$ interactions at beam momenta of 10,20,30 and 60~GeV/c.
\subsection{Acceptance criteria}
The MIPP calorimeter entrance plane is placed at a distance of 2458.6
cm from the center of the MIPP liquid hydrogen target volume. If the
neutral particle in the event impacts the calorimeter within a radius
of 75~cm from the beam axis, that neutral particle is considered
accepted for the purposes of this simulation. Similarly a slow proton
of momentum 0.206~GeV/c has a range of 10~cm in liquid hydrogen. if
the slow proton is more energetic than this, we accept the event. The
length in the beam direction of the target flask is 10.48~cm in length
along the beam direction. So on average, these protons should make it
out of the hydrogen and into the TPC. MIPP upgrade also plans to have
a recoil detector around the target, to detect wide angle slow protons.
\subsection{Calculated event rates}
We assume a spill of 4~seconds duration every 2~minutes. We assume a
DAQ rate of 3000 events per second during the spill. We assume the
machine is delivering beam 58\% of the time, the rest being devoted to
anti-proton stacking manipulations. These numbers are
conservative. Under these assumptions, MIPP should be able to log
5~millions events/day to disk. The events of interest will be among
these. The events where the hadron calorimeter has significant neutral
energy can be used to flag and filter these events for faster offline
processing.

In what follows, we calculate the events obtainable /day assuming the
total bandwidth is dedicated to the trigger in question. In practice,
we would select the charge of the beam and prescale the proton, kaon
and anti-proton beams as required.

\section{Results}

Table~\ref{cross} lists the inelastic cross sections of DPMJET
generated events for $pp,K^+p, K^-p$ and ${\bar p}p$ events and
compares them to known data obtained from PDG
listings. Table~\ref{mult} lists the mean charged multiplicities of
DPMJET generated events and compares them to data. The data
multiplicities were estimated by a fit of the form
$<n>=a+bln(E_a)+bln^2(E_a)$, where $E_a$ is the available energy in
the collision defined as $E_a=\sqrt s - m_{beam}-m_{target}$, and
a=2.45, b=0.32 and c=0.53. See ref~\cite{whit}. DPMJET does a
reasonable job at estimating the $pp$ total inelastic cross section
and mean multiplicity as a function of center of mass energy. It seems
to overstimate the $K^{\pm}p$ inelastic cross sections.
\begin{table}[tbh]
\begin{tabular}{|c|c|c|c|c|c|c|}
\hline
Beam Momentum & \multicolumn{2}{c}{pp inelastic} & \multicolumn{2}{|c}{K$^+$p inelastic} & \multicolumn{2}{|c|}{K$^-$p inelastic} \\
& \multicolumn{2}{c}{(mb)}& \multicolumn{2}{|c}{(mb)}& \multicolumn{2}{|c|}{(mb)}\\
\hline
GeV/c & DPMJET & data & DPMJET & data & DPMJET & data \\
\hline
10 & 32.29&30.0&21.05&13.96&22.09&19.96\\
\hline
20 &31.67&30.06&20.43&14.22&21.41&18.28\\
\hline
30 &31.55&30.89&20.24&15.33&21.13&18.28\\
\hline
60 &31.63&31.84&20.13&16.07&20.83&17.70\\
\hline
90 &31.84&30.50&-&16.12&-&17.29\\
\hline
\end{tabular}
\caption{Comparison of inelastic cross sections generated by DPMJET to data~\label{cross}}
\end{table}
\begin{table}[tb]
\caption{Comparison of mean charge multiplicities generated by DPMJET to data~\label{mult}}
\begin{center}
\begin{tabular}{|c|c|c|c|c|c|c|}
\hline
Beam Momentum & \multicolumn{2}{c}{pp inelastic} & \multicolumn{2}{|c}{K$^+$p inelastic} & \multicolumn{2}{|c|}{K$^-$p inelastic} \\
& \multicolumn{2}{c}{multiplicity} & \multicolumn{2}{|c}{multiplicity} & \multicolumn{2}{|c|}{multiplicity} \\
\hline
GeV/c & DPMJET & data & DPMJET & data & DPMJET & data \\
\hline
10 &3.70&3.27&3.97&3.45&3.59&3.45\\
\hline
20 &4.66&4.09&4.93&4.25&4.68&4.25\\
\hline
30 &5.27&4.62&5.52&4.77&5.32&4.77\\
\hline
60 &6.34&5.66&6.59&5.77&6.45&5.77\\
\hline
90 &7.04&6.31&-&6.42&-&6.42\\
\hline
\end{tabular}
\end{center}
\end{table}

\subsection{The reaction $pp\rightarrow n\pi^+p$} 
\begin{figure}[tbh]
\begin{center}
\includegraphics[width=\textwidth]{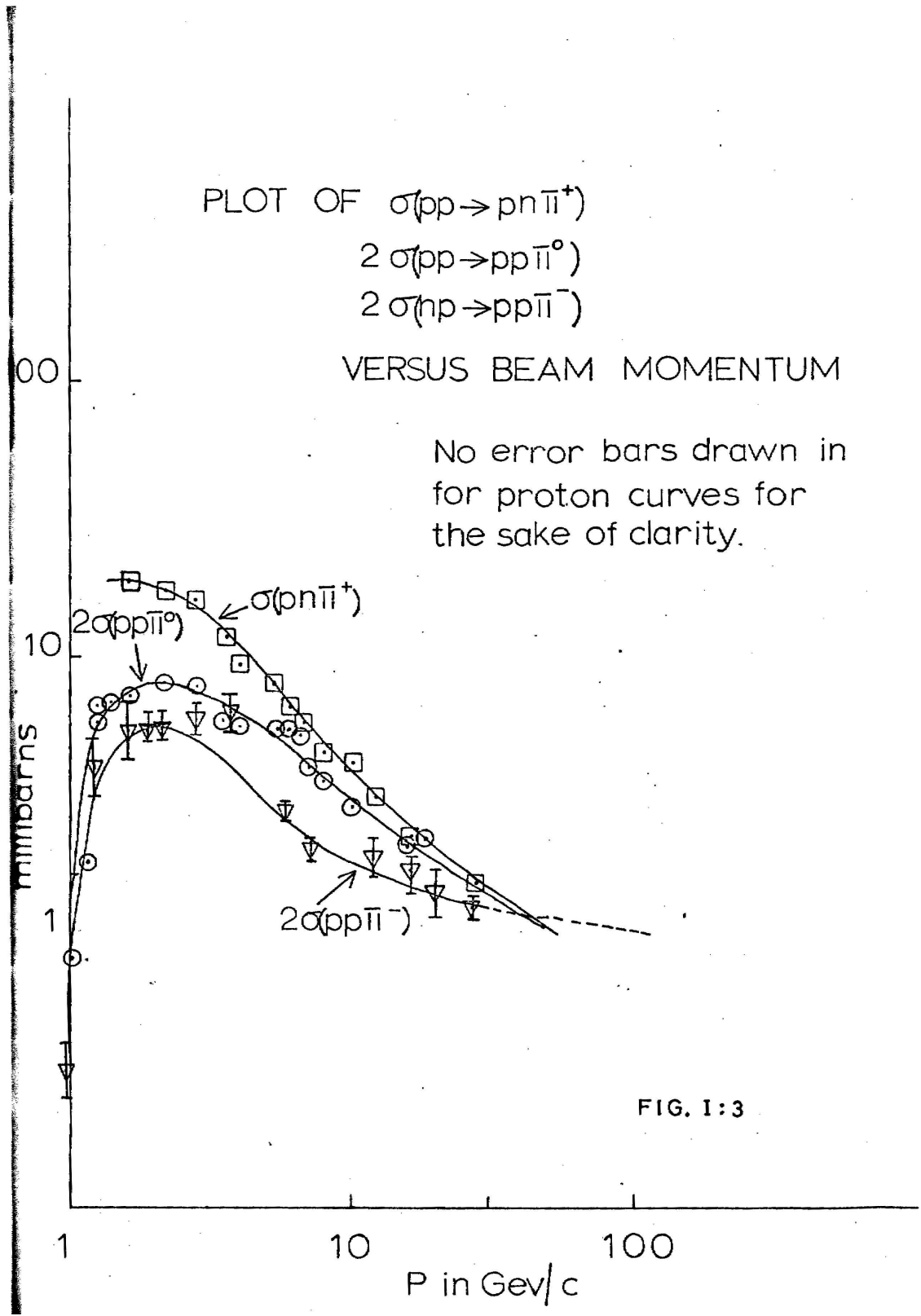}
\caption{The experimentally measured cross section 
for the reaction $pp\rightarrow pn\pi^+$ compared with similar diffractive processes.~\label{xsect1}}
\end{center}
\end{figure}

Figure~\ref{xsect1}
shows the cross section of $pp\rightarrow pn\pi^+$ events as a
function of beam momentum and compares them to the other diffractive
channels $np\rightarrow pp\pi^-$ and $pp\rightarrow pp\pi^0$. This
plot is taken from my thesis experiment~\cite{raja-thesis} which was
on the channel $np\rightarrow pp\pi^-$. Some of the cross section
points for the channel $np\rightarrow pp\pi^{-}$ were determined by a
technique that used ``tagged neutrons'' by observing $np$ elastic
scattering in the bubble chamber followed by the neutron
re-interacting in the chamber~\cite{ward-thesis}, which determines the 
neutron spectra.

\begin{figure}[tbh]
\begin{center}
\includegraphics[width=\textwidth]{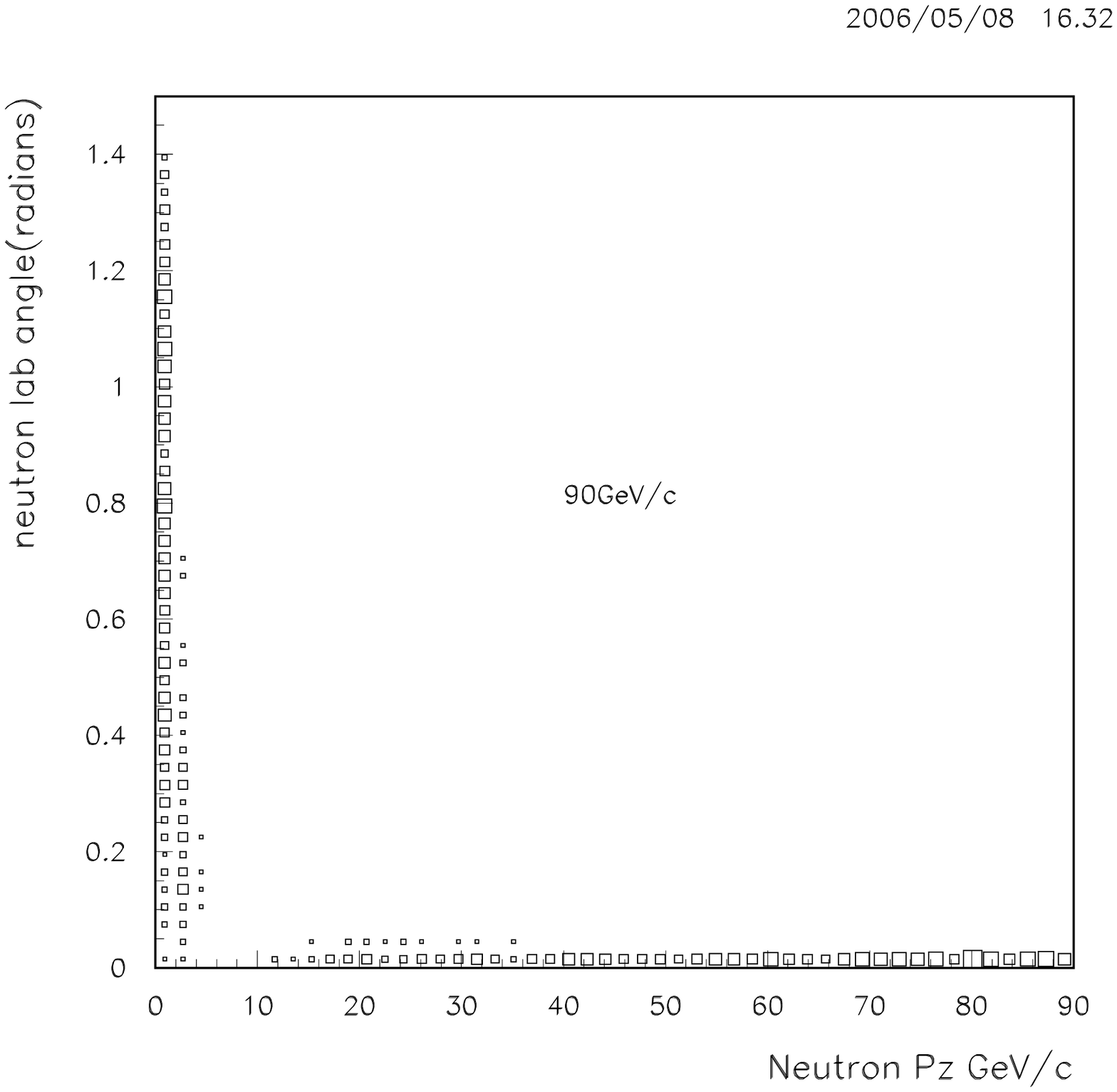}
\caption{The lab angle of the neutron as a function of the neutron momentum for the reaction $pp\rightarrow pn\pi^+$ for a beam momentum of 90~GeV/c. The wide angle neutrons are the result of target fragmentation.~\label{scat90}}
\end{center}
\end{figure}

Figure~\ref{scat90} shows the angle of the neutron for this reaction
as a function of the neutron momentum for an incident proton beam
momentum of 90~GeV/c. The wide angle neutrons are a result of target
fragmentation. The beam diffraction events have the neutron at a small
angle heading straight for the calorimeter, irrespective of the
neutron momentum. The acceptance of these neutrons is not increased
by bringing the calorimeter closer. However, the precision to which
the neutron 4-vector is known increases linearly as the calorimeter is
placed farther away from the interaction point, due to a better
measurement of the neutron angles.
Table~\ref{ppexp} tabulates the results of the DPMJET simulation. The
cross section for the process $pp\rightarrow pn\pi^+$ as calculated by
DPMJET is lower than the data by a factor of 3-4. This
graphically illustrates the problem of hadronic physics simulators. If
we use the data cross sections, we obtain 47,069 tagged neutrons in a
calorimeter with a 60~GeV/c proton beam. DPMJET evaluates this rate
as being 16,250. There are very few additional data points in this
channel since 1975, the time of my thesis. This fact alone 
accentuates our lack of knowledge of hadronic physics and the urgent
need for accurate high statistics data which the MIPP upgrade can
provide.
\begin{table}
\caption{Expected number of events/day using the DPMJET and data cross
sections for the process $pp\rightarrow pn\pi^+$. DPMJET underestimates
the cross section.~\label{ppexp}}
\begin{tabular}{|c|c|c|c|c|c|c|}
\hline
Beam Momentum & dpmjet & data  & dpmjet    & accepted & dpmjet & data \\
\hline
GeV/c         & mb      & mb   & generated & events   & events/day  & events/day \\
\hline
10 & 1.373 & 3.880 & 4252 & 135 & 6750 & 20532 \\
\hline
20 & 0.409 & 1.970 & 1290 & 207 & 10350 & 52581 \\
\hline
30 & 0.345 & 1.429 & 1092 & 314 & 15700 & 66511 \\
\hline
60 & 0.280 & 0.816 & 885  & 325 & 16250 & 47069 \\
\hline 
90 & 0.255 & 0.638 & 801  & 288 & 14400 & 37600 \\
\hline
\end{tabular}
\end{table}
\begin{figure}[tbh]
\begin{center}
\includegraphics[width=\textwidth]{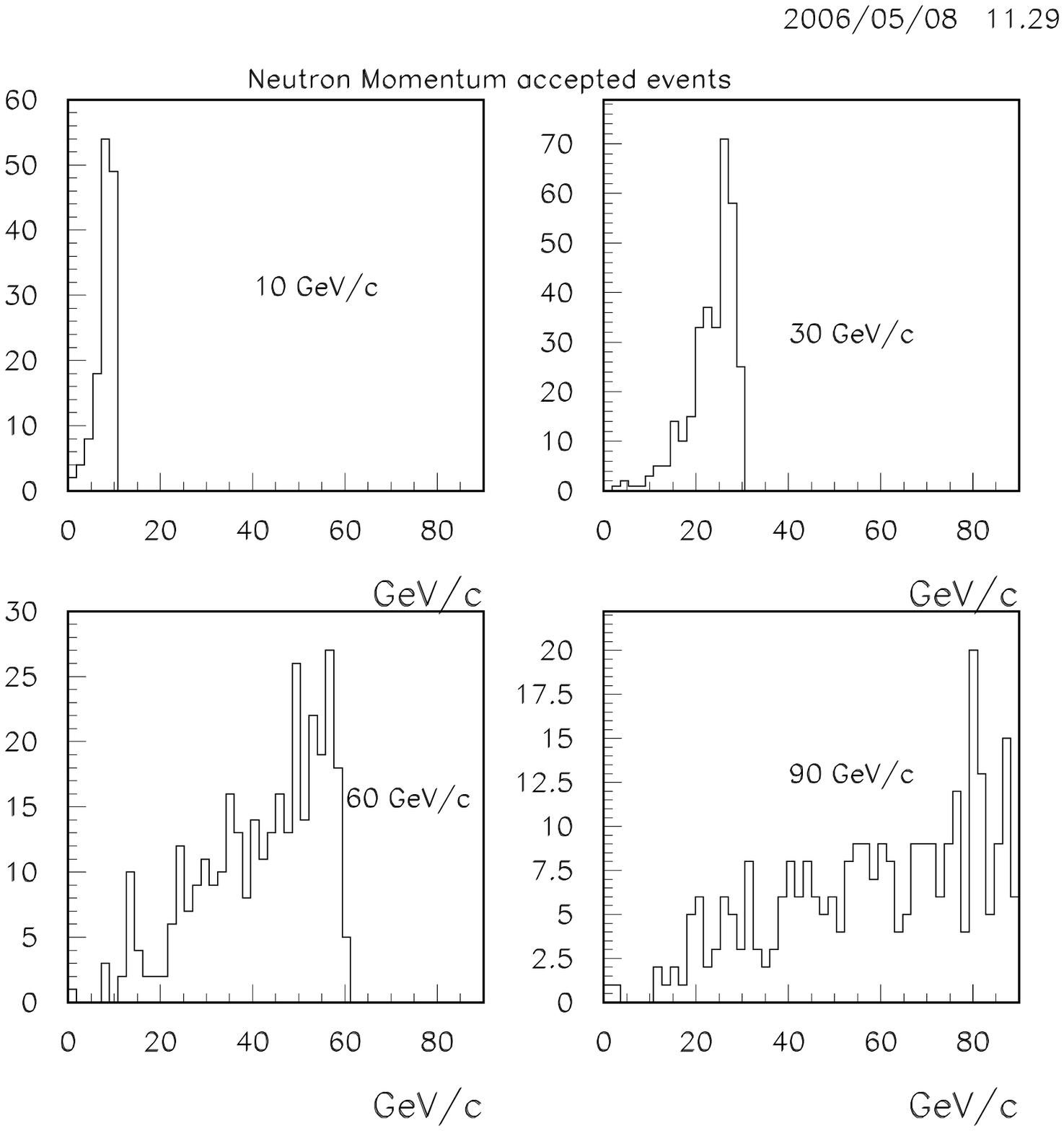}
\caption{Momentum spectrum  of accepted neutrons for incident proton momenta of
10~GeV/c, 30~GeV/c, 60~GeV/c and 90~GeV/c 
for the process $pp\rightarrow pn\pi^+$.~\label{pneut}}
\end{center}
\end{figure}
\begin{figure}[tbh]
\begin{center}
\includegraphics[width=\textwidth]{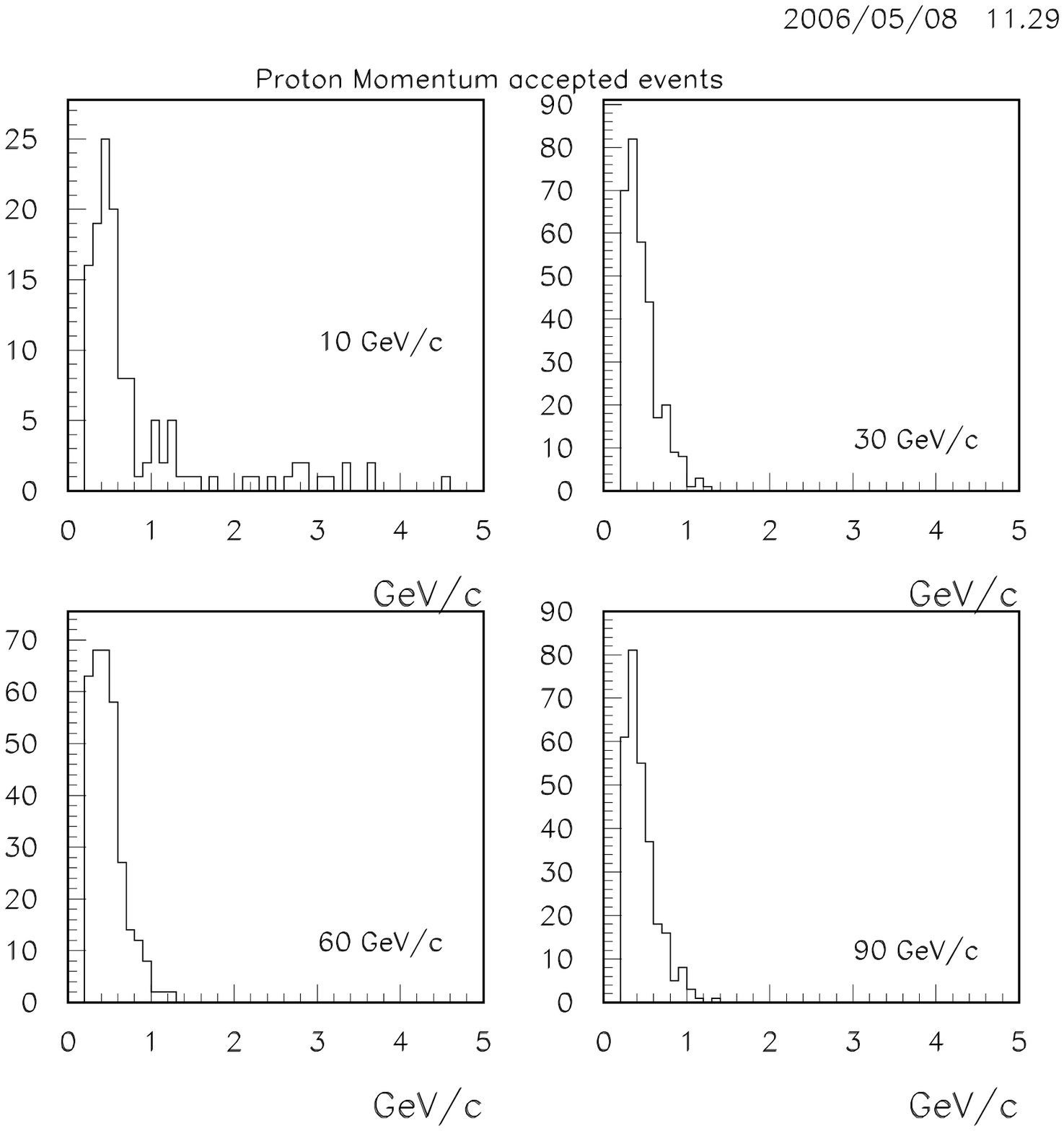}
\caption{Momentum spectrum  of accepted protons for incident proton momenta of
10~GeV/c, 30~GeV/c, 60~GeV/c and 90~GeV/c
for the process $pp\rightarrow pn\pi^+$.~\label{pprot}}
\end{center}
\end{figure}
\begin{figure}[tbh]
\begin{center}
\includegraphics[width=\textwidth]{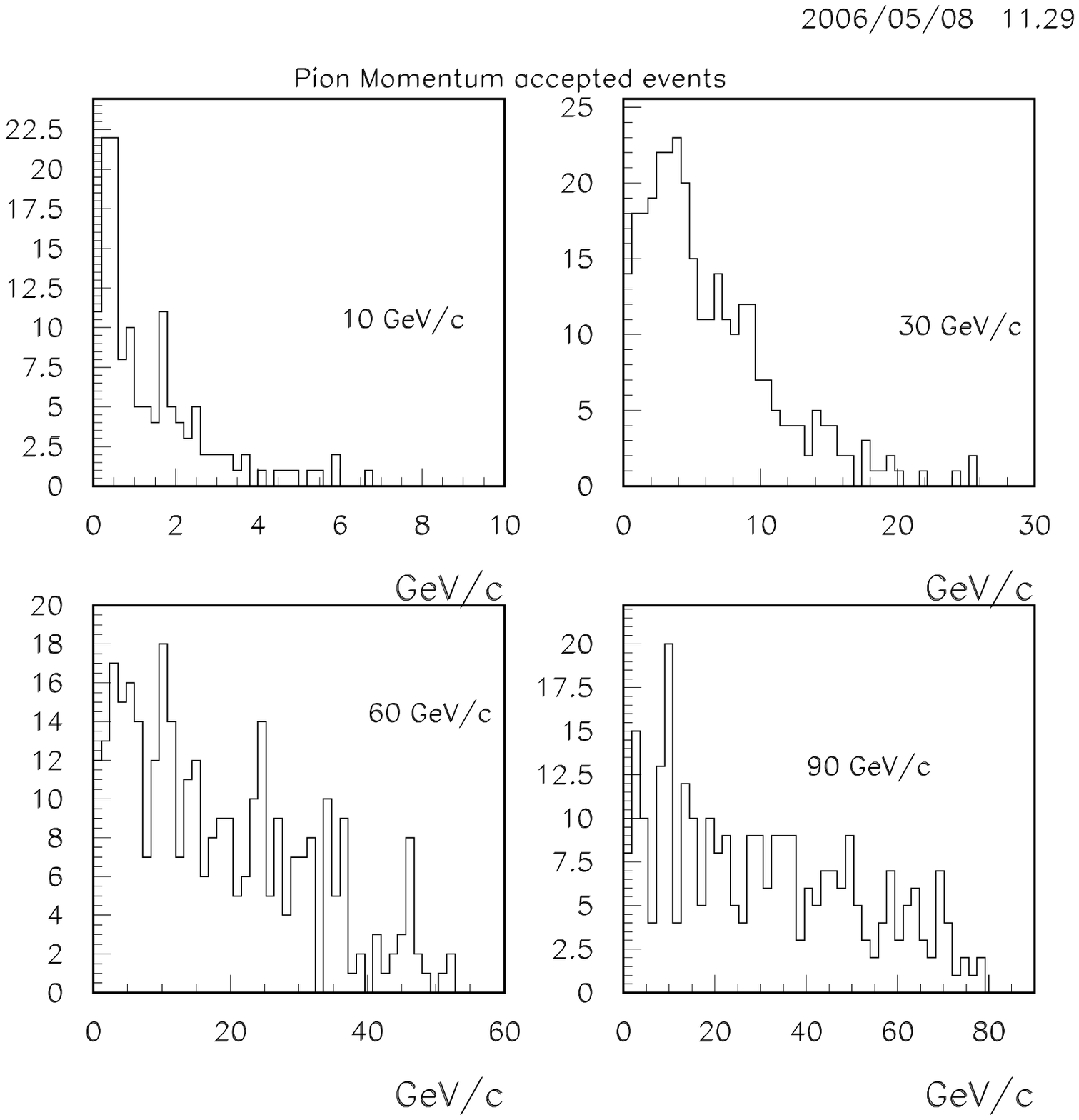}
\caption{Momentum spectrum  of accepted $\pi^+$ particles for incident 
proton momenta of
10~GeV/c, 30~GeV/c, 60~GeV/c and 90~GeV/c
for the process $pp\rightarrow pn\pi^+$.~\label{ppion}}
\end{center}
\end{figure}
Figures~\ref{pneut},~\ref{pprot} 
and ~\ref{ppion} show the neutron, proton and pion spectra of accepted events as a 
function of beam momentum.

\subsection{The reaction $K^+p\rightarrow K^0_L\pi^+p$} 
\begin{figure}[tbh]
\begin{center}
\includegraphics[width=\textwidth]{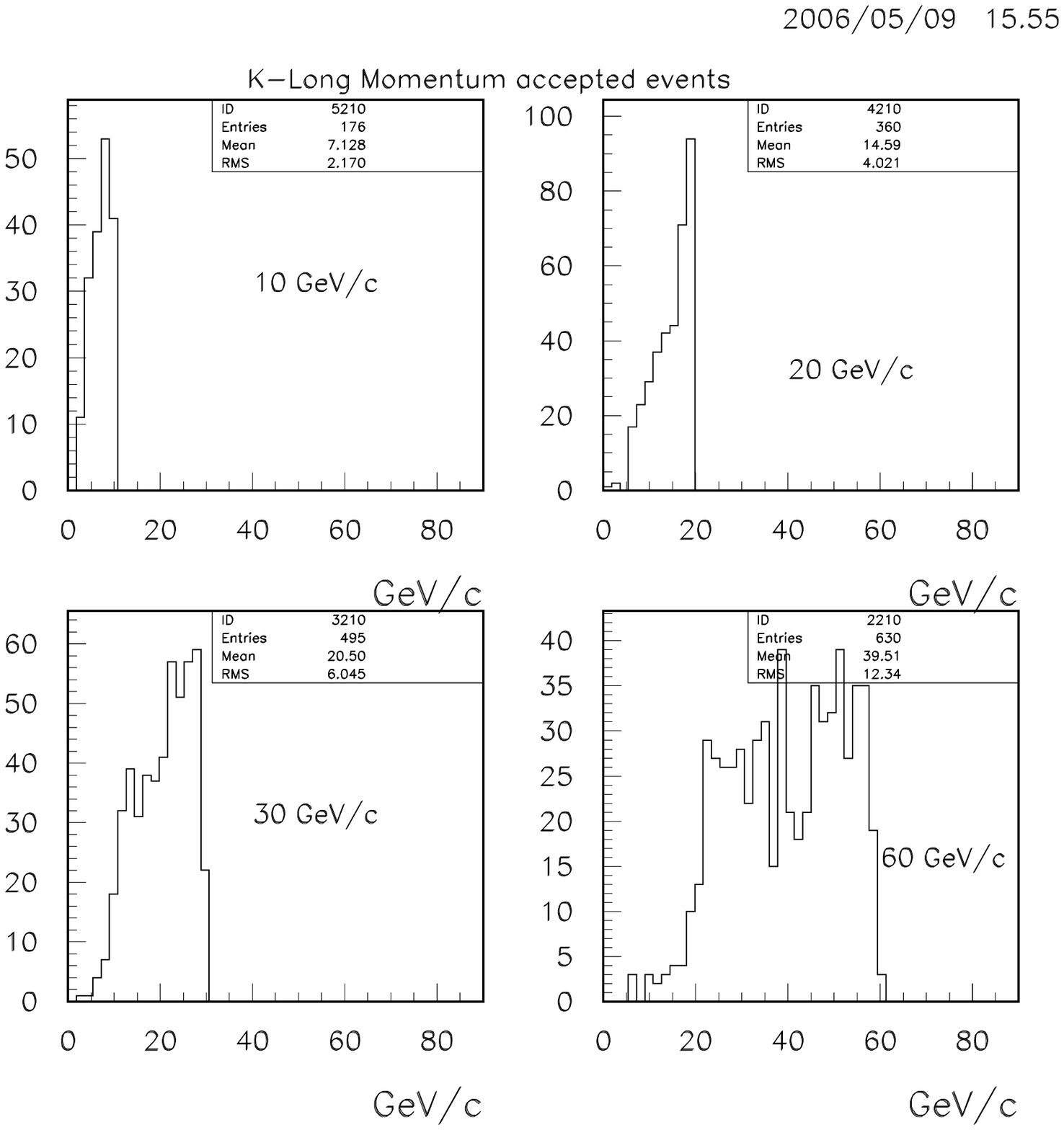}
\caption{Momentum spectrum  of accepted $K^0_L$ particles
 for incident $K^+$ momenta of
10~GeV/c, 20~GeV/c, 30~GeV/c and 60~GeV/c
for the process $K^+p\rightarrow pK^0_L\pi^+$.~\label{klong}}
\end{center}
\end{figure}

\begin{figure}[tbh]
\begin{center}
\includegraphics[width=\textwidth]{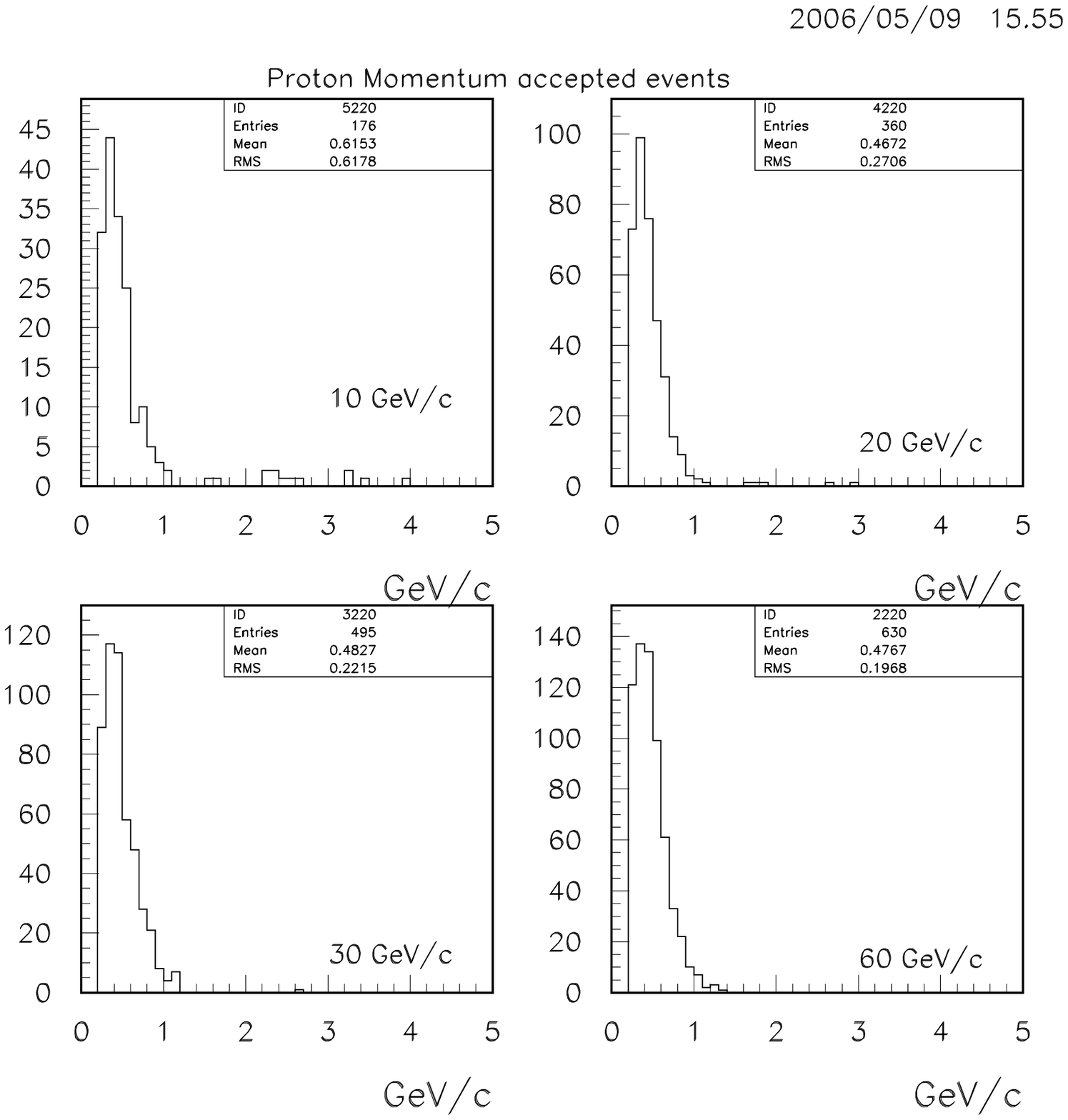}
\caption{Momentum spectrum  of accepted protons for incident $K^+$ momenta of
10~GeV/c, 20~GeV/c, 30~GeV/c and 60~GeV/c 
for the process $K^+p\rightarrow pK^0_L\pi^+$.~\label{kprot}}
\end{center}
\end{figure}

\begin{figure}[tbh]
\begin{center}
\includegraphics[width=\textwidth]{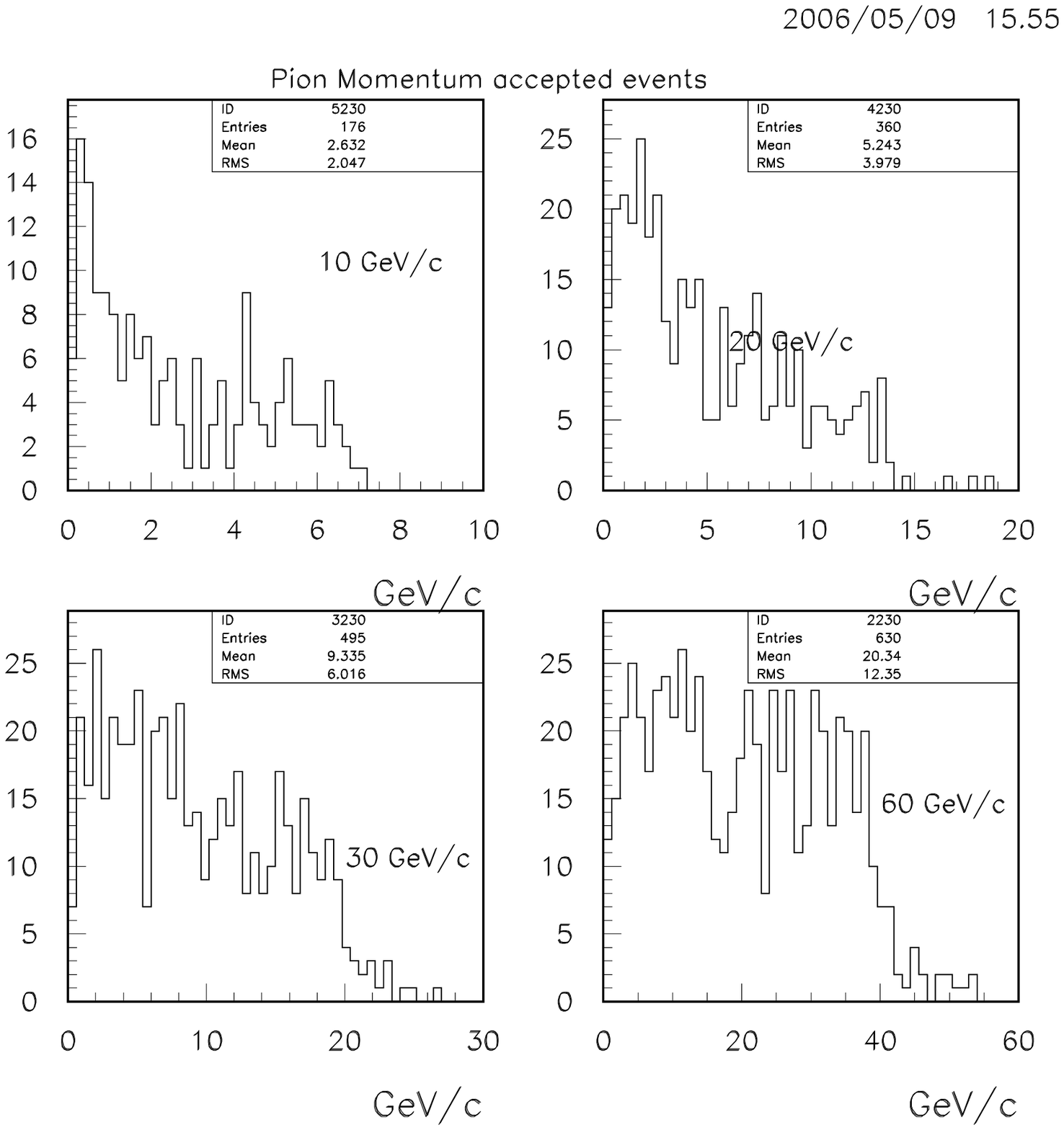}
\caption{Momentum spectrum  of accepted $\pi^+$ particles for incident 
$K^+$ momenta of
10~GeV/c, 20~GeV/c, 30~GeV/c and 60~GeV/c 
for the process $K^+p\rightarrow pK^0_L\pi^+$.~\label{kpion}}
\end{center}
\end{figure}

Table~\ref{kpexp} tabulates the DPMJET cross section for the process
$K^+p\rightarrow pK^0_L\pi^+$. We have not attempted to compare the
DPMJET cross sections to data cross sections, though the agreement in
this channel may be better than in the $pp$
case. Figures~\ref{klong},~\ref{kprot} and ~\ref{kpion} show the
momentum spectra of $K^0_L$, proton and pions in the accepted events,
as a function of beam momenta.
\begin{table}[tbh]
\caption{Expected number of events/day using the DPMJET 
for the process $K^+p\rightarrow pK^0_L\pi^+$. ~\label{kpexp}}
\begin{tabular}{|c|c|c|c|c|c|}
\hline
Beam Momentum & dpmjet & dpmjet  & dpmjet    & accepted & dpmjet  \\
\hline
GeV/c         & inel. mb & mb   & generated  & events   & events/day  \\
\hline
10 & 21.05 & 0.226 & 2152 & 176 & 4400 \\
\hline
20 & 20.43 & 0.113 & 1110 & 360 & 9000 \\
\hline
30 & 20.24 & 0.093 & 923 & 495 & 12375 \\
\hline
60 & 20.13 & 0.086 & 855  & 630 & 15750 \\
\hline 
\end{tabular}
\end{table}

\subsection{The reaction $K^-p\rightarrow K^0_L\pi^-p$} 
\begin{figure}[tbh]
\begin{center}
\includegraphics[width=\textwidth]{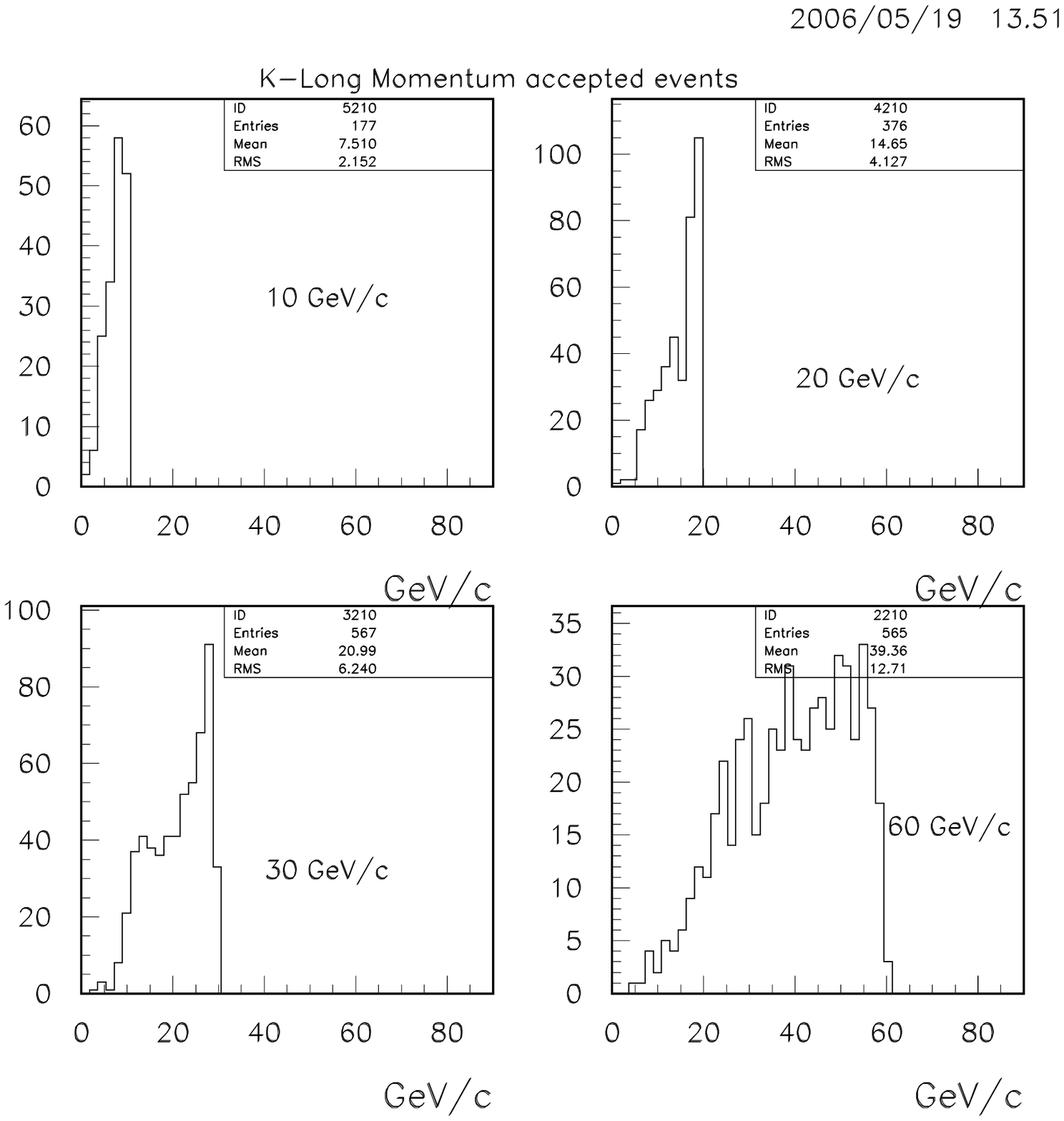}
\caption{Momentum spectrum  of accepted $K^0_L$ particles
 for incident $K^-$ momenta of
10~GeV/c, 20~GeV/c, 30~GeV/c and 60~GeV/c
for the process $K^-p\rightarrow pK^0_L\pi^-$.~\label{klongm}}
\end{center}
\end{figure}

\begin{figure}[tbh]
\begin{center}
\includegraphics[width=\textwidth]{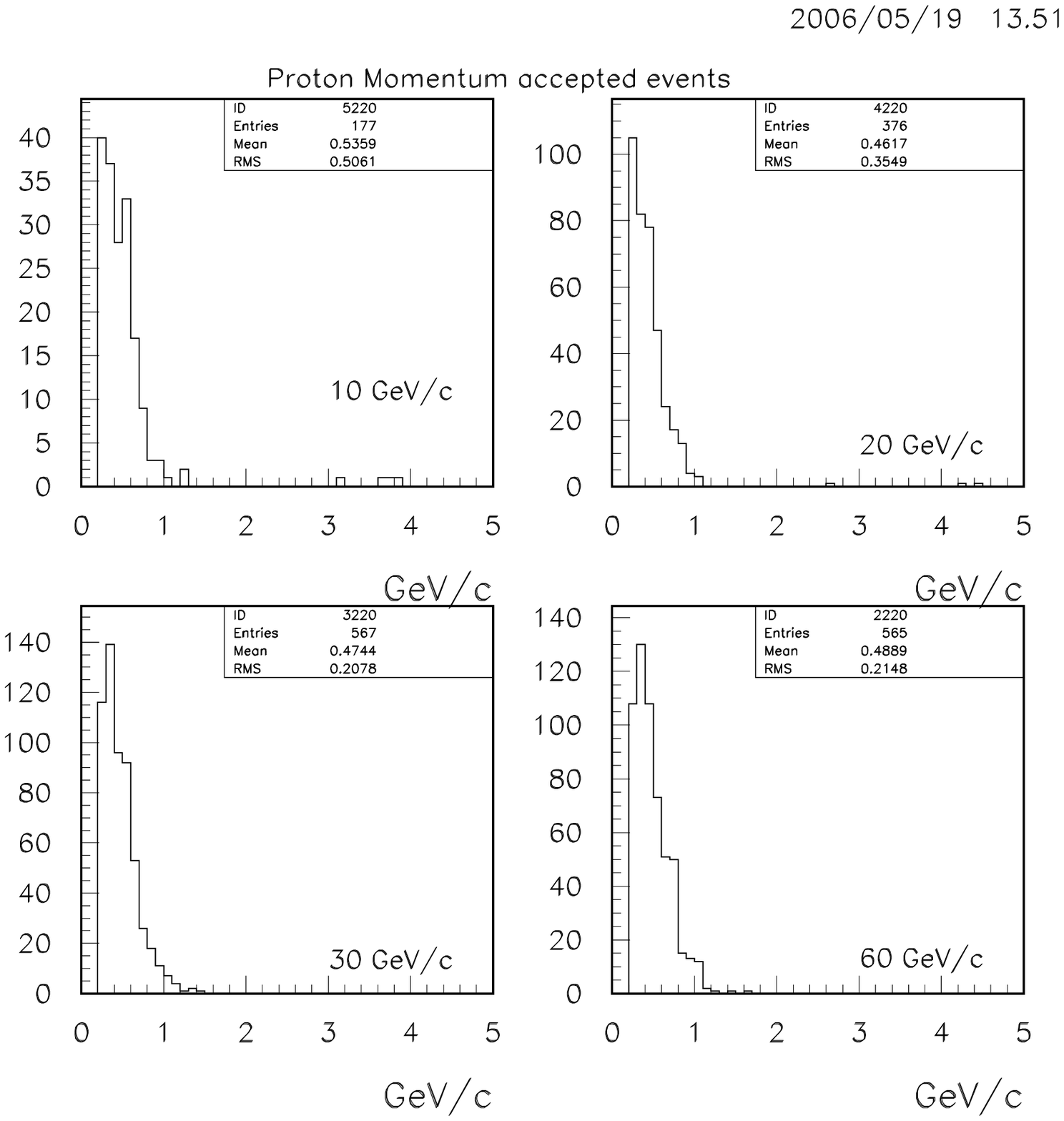}
\caption{Momentum spectrum  of accepted protons for incident $K^-$ momenta of
10~GeV/c, 20~GeV/c, 30~GeV/c and 60~GeV/c 
for the process $K^-p\rightarrow pK^0_L\pi^-$.~\label{kprotm}}
\end{center}
\end{figure}

\begin{figure}[tbh]
\begin{center}
\includegraphics[width=\textwidth]{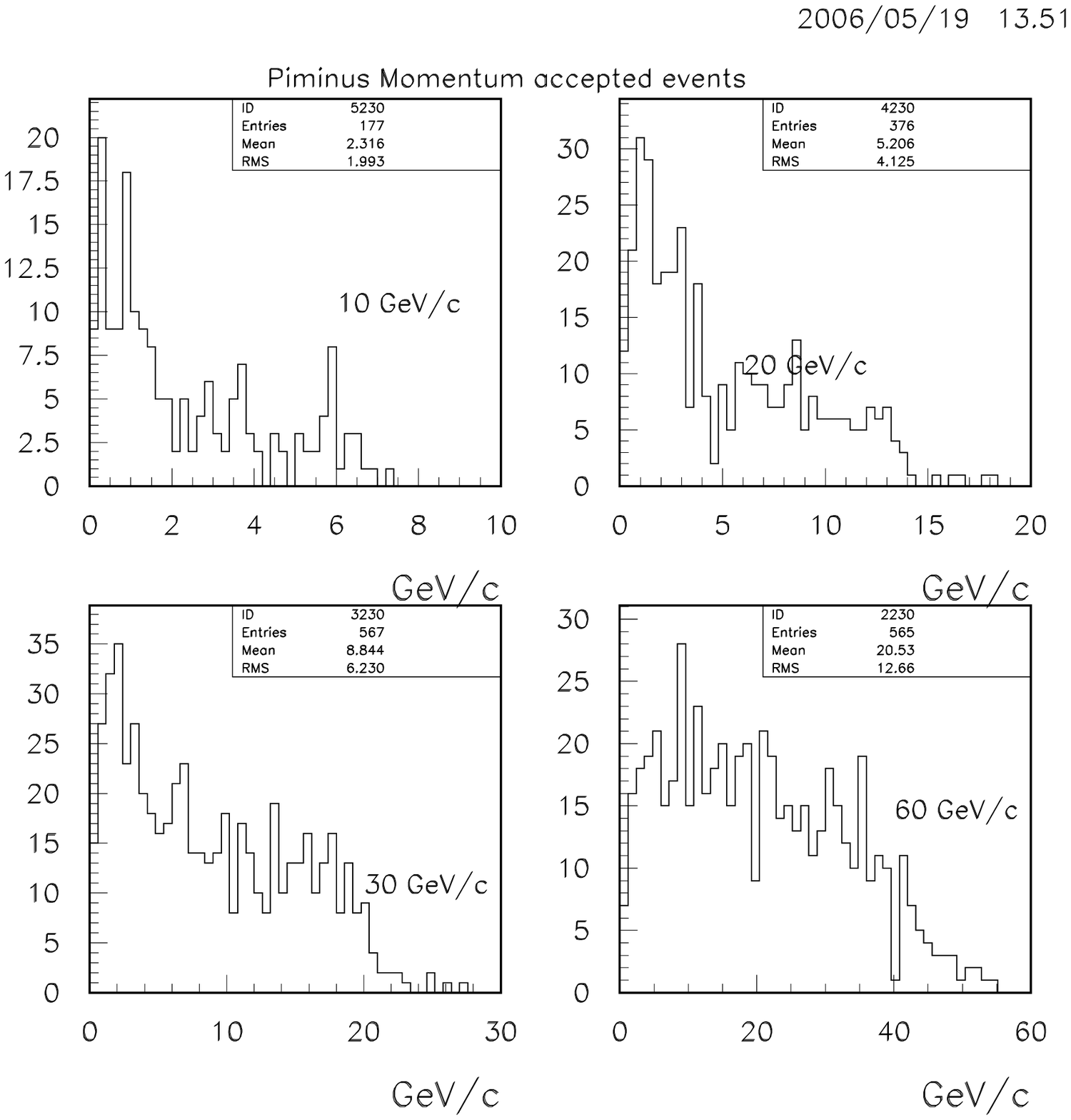}
\caption{Momentum spectrum  of accepted $\pi^-$ particles for incident 
$K^-$ momenta of
10~GeV/c, 20~GeV/c, 30~GeV/c and 60~GeV/c 
for the process $K^-p\rightarrow pK^0_L\pi^-$.~\label{kpionm}}
\end{center}
\end{figure}

Table~\ref{kmexp} tabulates the DPMJET cross section for the process
$K^-p\rightarrow pK^0_L\pi^-$. We have again not attempted to compare the
DPMJET cross sections to data cross sections, though the agreement in
this channel may be better than in the $pp$
case. Figures~\ref{klongm},~\ref{kprotm} and ~\ref{kpionm} show the
momentum spectra of $K^0_L$, proton and pions in the accepted events,
as a function of beam momenta.

\begin{table}[tbh]
\caption{Expected number of events/day using the DPMJET 
for the process $K^-p\rightarrow pK^0_L\pi^-$. ~\label{kmexp}}
\begin{tabular}{|c|c|c|c|c|c|}
\hline
Beam Momentum & dpmjet & dpmjet  & dpmjet    & accepted & dpmjet  \\
\hline
GeV/c         & inel mb & mb   & generated  & events   & events/day  \\
\hline
10 & 22.09 & 0.178 & 1616 & 177 & 4425\\
\hline
20 & 21.41 & 0.120 & 1122 & 376 & 9400 \\
\hline
30 & 21.13 & 0.105 & 993 & 567 & 14175 \\
\hline
60 & 20.83 & 0.082 & 784 & 565 & 14125\\
\hline 
\end{tabular}
\end{table}
\subsection{The reaction ${\bar p}p\rightarrow {\bar n}\pi^-p$} 
Table~\ref{apexp} tabulates the results of the DPMJET simulation. We
expect the same mismatch between the DPMJET cross sections and the
data in this channel as in the $pp$ case, though data are even
sparser. We estimate the number of anti-neutrons expected as a
function of beam momentum, with the proviso that the data may have
higher rates by a factor of 3-4.
\begin{table}[tbh]
\caption{Expected number of events/day using the DPMJET 
for the process $\bar{p}p\rightarrow p\bar{n}\pi^-$. ~\label{apexp}}
\begin{tabular}{|c|c|c|c|c|c|}
\hline
Beam Momentum & dpmjet & dpmjet  & dpmjet    & accepted & dpmjet  \\
\hline
GeV/c         & inel. mb & mb   & generated  & events   & events/day  \\
\hline
10 & 40.63 & 0.223 & 1097 & 133 & 6650 \\
\hline
20 & 38.08 & 0.108 & 568 & 229 & 11450 \\
\hline
30 & 36.96 & 0.099 & 536 & 270 & 13500 \\
\hline
60 & 35.60 & 0.068 & 381 & 271 & 13550 \\
\hline 
\end{tabular}
\end{table}
\begin{figure}[tbh]
\begin{center}
\includegraphics[width=\textwidth]{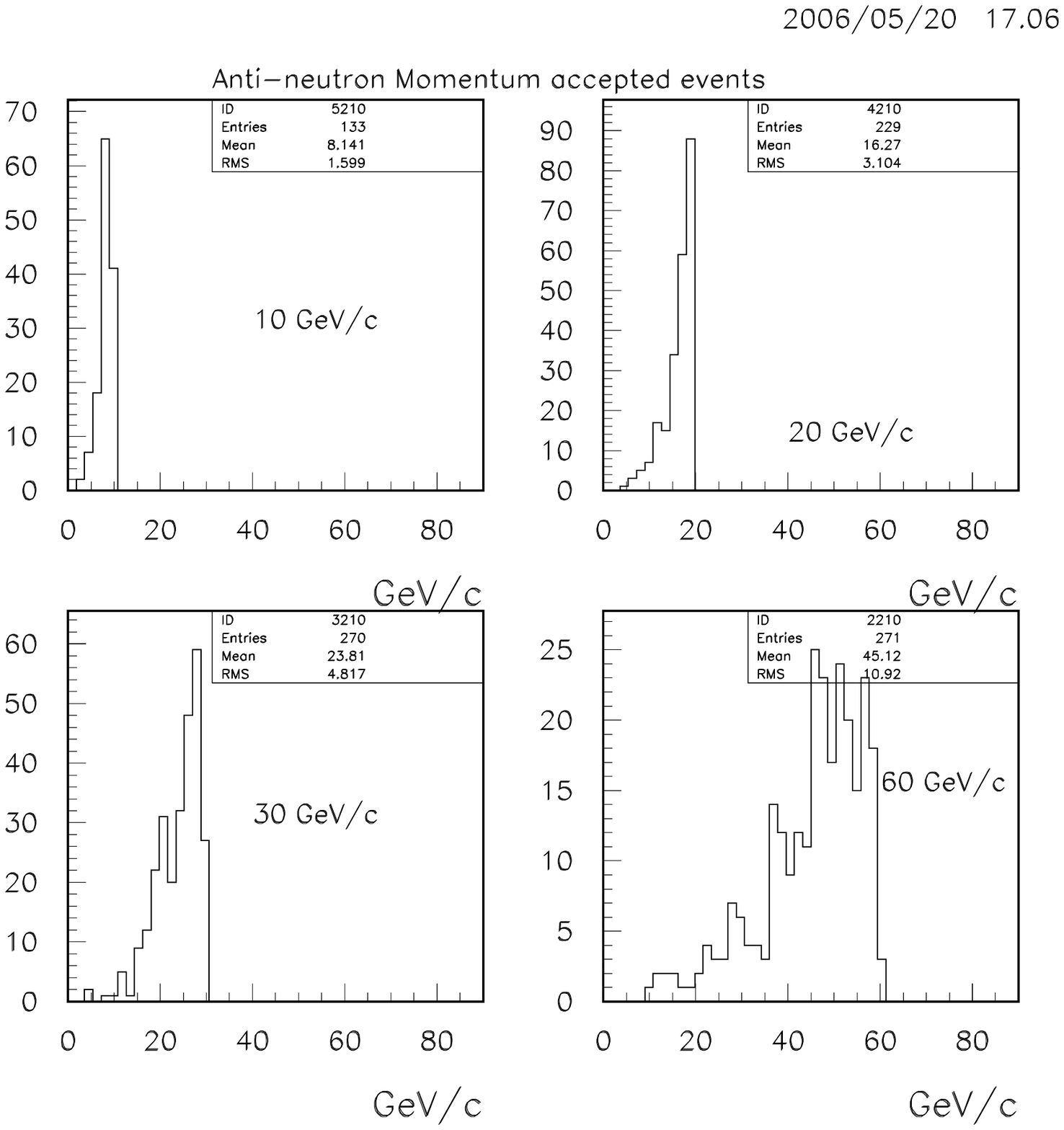}
\caption{Momentum spectrum  of accepted anti-neutrons for incident anti-proton momenta of
10~GeV/c, 20~GeV/c, 30~GeV/c and 60~GeV/c 
for the process $\bar{p}p\rightarrow p\bar{n}\pi^-$.~\label{apneut}}
\end{center}
\end{figure}
\begin{figure}[tbh]
\begin{center}
\includegraphics[width=\textwidth]{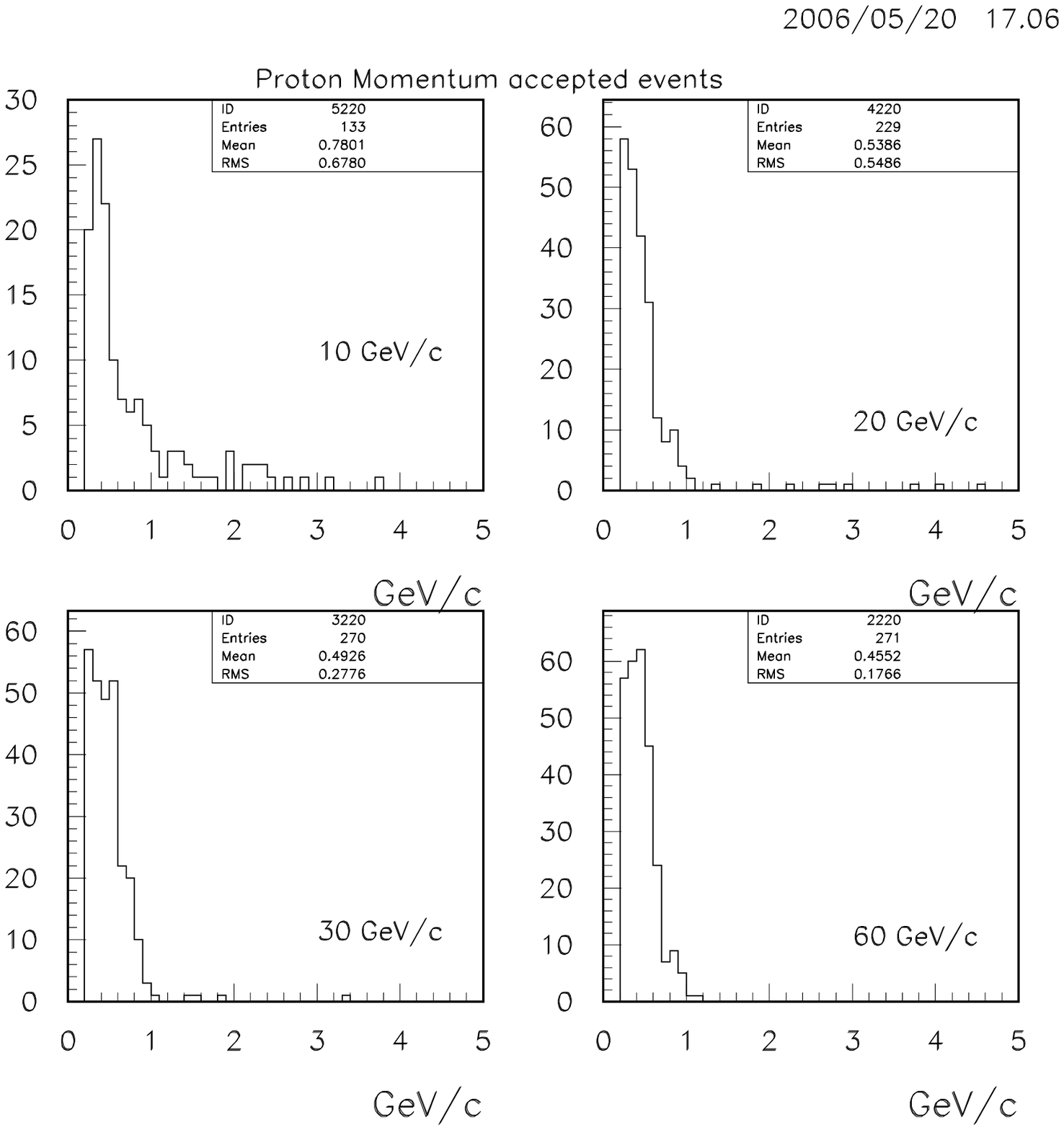}
\caption{Momentum spectrum  of accepted protons for incident anti-proton momenta of
10~GeV/c, 20~GeV/c, 30~GeV/c and 60~GeV/c
for the process $\bar{p}p\rightarrow p\bar{n}\pi^-$.~\label{approt}}
\end{center}
\end{figure}
\begin{figure}[tbh]
\begin{center}
\includegraphics[width=\textwidth]{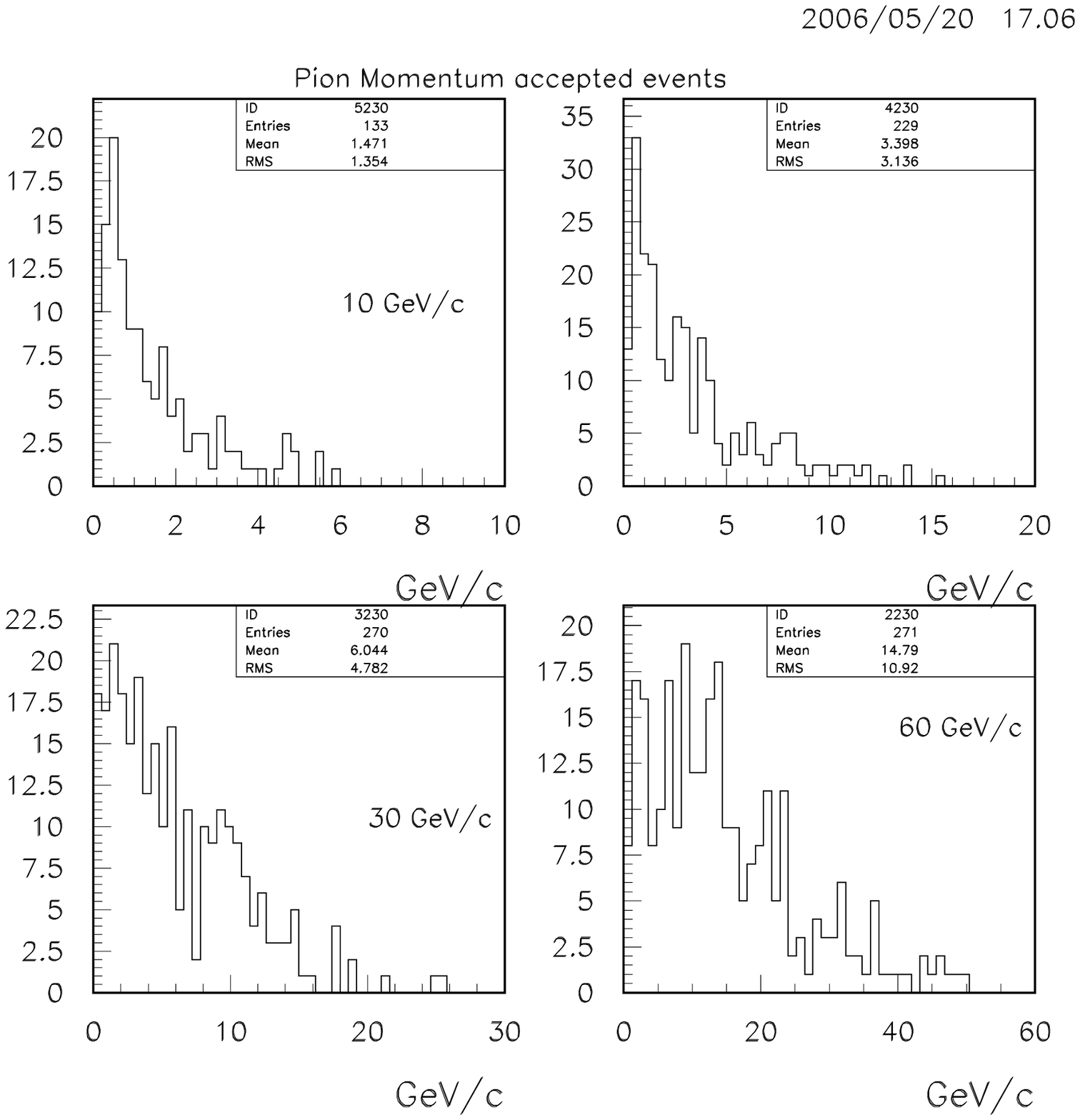}
\caption{Momentum spectrum  of accepted $\pi^-$ particles for incident 
anti-proton momenta of
10~GeV/c, 20~GeV/c, 30~GeV/c and 60~GeV/c
for the process $\bar{p}p\rightarrow p\bar{n}\pi^-$.~\label{appion}}
\end{center}
\end{figure}
Figures~\ref{apneut},~\ref{approt} 
and ~\ref{appion} show the anti-neutron, proton and pion spectra of accepted events as a 
function of beam momentum.

\section{Conclusions}

We propose a scheme by which we obtain tagged neutron, anti-neutron
and $K^0_L$ beams using an upgraded MIPP spectrometer. A test
calorimeter placed behind the RICH counter in MIPP will enable the
study of neutral particle response in the calorimeter. The momenta of
the tagged neutral particle will be known on a particle by particle
basis to better than 2\%. The upgraded MIPP calorimeter will also
permit the acquisition of hadroproduction data on a number of nuclei
of unprecedented quality and precision. Such data will dramatically
improve our knowledge of QCD processes and our ability to simulate
hadronic showers.
We note in passing that the kaon tagging results in pure $K^0$ and
$\bar{K^0}$ states at the production vertex. The physics implications
of this for possible CP-violation studies need to be investigated
further. The author wishes to acknowledge useful conversations with
Marcel Demarteau and Nickolas Solomey.

\end{document}